\documentstyle[preprint,aps]{revtex}

\draft
\preprint{cond-mat/9405039, ETH-TH/94-15}
\title{ %
Superconductivity in the Two-Dimensional $t$-$J$ Model at Low Hole
Doping
}

\author{E. S. Heeb and T. M. Rice %
\address{Theoretische Physik, Eidgen\"ossische Technische Hochschule
H\"onggerberg, CH-8093 Z\"urich, Switzerland}}

\begin{document}
\maketitle

\begin{abstract}
By combining a generalized Lanczos scheme with the variational Monte
Carlo method we can optimize the short- and long-range properties of
the groundstate separately.  This allows us to measure the long-range
order of the groundstate of the $t$-$J$ model as a function of the
coupling constant $J/t$, and identify a region of finite d-wave
superconducting long-range order.  With a lattice size of 50 sites we
can reliably examine hole densities down to 0.16.
\end{abstract}

\pacs{PACS numbers: 71.10.+x, 71.27.+a, 74.20.Mn}

\narrowtext

Presently one of the most interesting questions in the study of
strongly correlated electron systems is to determine the region of
superconductivity in the phase diagram of the two-dimensional $t$-$J$
model\cite{Dagotto-Review,tJ-Hubbard}.  Despite considerable effort,
high temperature expansion\cite{putikka-etal92,high-temp-various} and
Quantum Monte Carlo
calculations\cite{sorella-etal89,boninsegni-manousakis} are unable to
provide conclusive evidence for or against superconductivity.
Although variational calculations\cite{tJ-VMC} are in favor of
superconductivity, they are not able to establish independent evidence
in a region of the phase diagram where magnetically ordered phases are
competing with superconductivity and have almost the same energy.  To
date the strongest unbiased indication for superconducting order in
the groundstate comes from exact diagonalization of a $4\times 4$
cluster\cite{dagotto-riera93}.  With such a small cluster only a few
filling fractions are available and it remains open how much the
finite size effects contribute to the results.  In this letter we will
present results on $\sqrt{50}\times\sqrt{50}$ cluster, which allows us
to reach smaller hole dopings and reduce the finite size effects.  By
varying the long-range and short-range correlations separately we can
go beyond a standard variational approach and we are able to measure
the long-range pair-pair correlation function.  This enables us to
identify a region of a d-wave superconducting groundstate in the phase
diagram.

The $t$-$J$ model has attracted considerable attention, because it is
the simplest model to describe strongly correlated electrons.  The
perturbative approaches are not possible since there is no exact
solution that could be used as a starting point.  In contrast, many of
the numerical methods that have been used for strongly correlated
electrons do not rely on a small parameter in the Hamiltonian.  Exact
diagonalization provides reliable results of a variety of properties.
However, for the two-dimensional $t$-$J$ model only systems up to a
size of $4 \times 4$ have been investigated for all
fillings\cite{dagotto-etal92}.  The results for bigger systems are
restricted to certain fillings only, {\it e.g.}, 2 holes on 26
sites\cite{poilblanc93}.  The limited number of systems and fillings
makes it difficult, if not impossible, to obtain reliable information
about the thermodynamic limit.  Properties that are related to the
short range behavior like the energy ({\it e.g.}, total energy,
binding energy, spectral function) are more reliable than those
involving long-range behavior.  Indeed, the evidence for the presence
or absence of superconducting long-range order from exact
diagonalization is very limited\cite{dagotto-riera93,ohta-etal}.
High-temperature expansions directly lead to results which are valid
in the thermodynamic limit\cite{putikka-etal92}, but the extrapolation
from finite temperatures to $T = 0$ is difficult.  So far, only the
equal time behavior of the spin and charge degrees of freedom have
been successfully analyzed\cite{high-temp-various}, while the question
of superconducting order remains open.  The Quantum Monte Carlo
methods\cite{QMC} which are very powerful for weakly interacting
systems show severe restrictions due to the fermion sign problem
especially in the strong coupling limit.  The available results cover
mainly the intermediate coupling regime of the Hubbard
model\cite{sorella-etal89} whereas for the inherently strong coupling
$t$-$J$ model only the cases of one and two holes have been
considered\cite{boninsegni-manousakis}.

With variational approaches no fermion sign problem occurs and the
systems are big enough to show only small finite size effects.  If
enough information about the symmetry of the groundstate is known, a
variational wavefunction can be constructed to model this groundstate.
Typically, such a variational wavefunction is given analytically from
a mean field ansatz, so that even for fermions expectation values can
be readily evaluated by Monte Carlo sampling\cite{ceperley77}.  It is
well known that in the $t$-$J$ model various phases with different
broken symmetries compete with each other.  The energies of these
phases will be close to each other at small dopings so that the energy
differences are comparable to the error introduced by using modified
mean field forms for the variational wavefunctions.  The results of
these variational studies remain therefore inconclusive to some
extent, and the regions of stability of the various phases in the
phase diagram can only be estimated qualitatively\cite{tJ-VMC}.

Systematic iterative improvements have been used to remove the bias in
the choice of the wavefunction.  These methods range from the power
method\cite{chen-lee-PM} to Lanczos iterations \cite{Lanczos-VMC}.
While they do remove the bias, these methods are restricted to a few
iterations only.  The computing time needed to reduce the statistical
error increases rapidly with the number of iterations.  In this letter
we will use a generalized Lanczos approach which optimizes the short-
and long-range correlations separately and extract the relevant
information from one iteration.

The Hamiltonian for the $t$-$J$ model is defined in the subspace with
no doubly occupied sites as
\begin{equation}
H = -t \sum_{<i,j>,\sigma}
            \left(\tilde{c}^\dagger_{i,\sigma}
                  \tilde{c}_{j,\sigma} + {\rm h.c.}\right)
    + J \sum_{<i,j>}\left({\bf S}_i \cdot {\bf S}_j -
                          {1 \over 4} n_i n_j \right)
\end{equation}
where
$\tilde{c}^\dagger_{i,\sigma} %
= c^\dagger_{i,\sigma}\left(1 - n_{i,-\sigma}\right)$
prevent doubly occupied sites and the rest of the notation is
standard.  We perform our calculations on a 50 sites cluster with
periodic boundary conditions and periods $(7,1)$ and $(-1,7)$.  Exact
diagonalization\cite{dagotto-riera93} has found that at the largest
distance available in the $4\times4$ cluster the pair-pair correlation
function
$C(R) = (1/N) \sum_i{<\Delta^\dagger_i \Delta_{i+R}>}$
corresponding to a nearest-neighbor singlet d-wave pairing
operator
$\Delta_i = (1/2) %
\sum_\sigma{c_{i,-\sigma}\left(%
  c_{i+\hat{x},\sigma} + c_{i-\hat{x},\sigma} %
- c_{i+\hat{y},\sigma} - c_{i-\hat{y},\sigma}%
\right)}$
is dominant.  This long-range order is realized in a mean field
wavefunction as follows
\begin{equation}
\left|\Psi\left(D,\mu\right)\right> = %
{\cal P}_{\rm G} {\cal P}_{\rm N}\prod_{\bf k}%
\left(%
u_{\bf k} + v_{\bf k} %
c^\dagger_{{\bf k},\uparrow} %
c^\dagger_{-{\bf k},\downarrow} %
\right) %
\left|0\right>
\label{wavefunction}
\end{equation}
where ${\cal P}_{\rm G}$ and ${\cal P}_{\rm N}$ are the Gutzwiller- and
N-particle projectors, respectively.  The ratio
$v_{\bf k}/u_{\bf k} = \Delta_{\bf k}/\left(\xi_{\bf k} +%
\sqrt{\xi^2_{\bf k} + \Delta^2_{\bf k}}\right)$ with
$\xi_{\bf k} = -2\left(%
\cos\left({\bf k}_x\right) + \cos\left({\bf k}_y\right)%
\right) - \mu$
has the standard BCS form.  With
$\Delta_{\bf k} = D\left(%
\cos\left({\bf k}_x\right) - \cos\left({\bf k}_y\right)%
\right)$
this wavefunction has by construction a finite long-range d-wave order
in the thermodynamic limit for a finite $D$-parameter, whereas for $D
= 0$ it reduces to the Gutzwiller projected Fermi sea.  For the ${\bf
k}$ points where $\Delta_{\bf k}$ has a node the ratio $v_{\bf
k}/u_{\bf k}$ is not well defined when $\xi_{\bf k} < 0$.  In the
thermodynamic limit the nodes of $\Delta_{\bf k}$ in the Brillouin
zone are negligible.  The effect of these nodes on the wavefunction
accounts for much of the finite size effects.  Due to the tilted
periodic boundary conditions the 50 sites lattice has only one point
with $\Delta_{\bf k} = 0$ (at ${\bf k} = 0$) and is thus an optimal
choice to reduce the finite size effects.  Since ${\bf k} = 0$ is deep
inside the Fermi sea we set $v_{{\bf k}=0} \to 1$ and $u_{{\bf k}=0}
\to 0$ which leads to a ratio $v_{\bf k}/u_{\bf k} \to \infty$.  In an
actual calculation we choose a large but finite ratio.  For $D
\to 0$ the choice of this ratio has a bigger influence on the
wavefunction and the Fermi sea will be defined as the extrapolation
from small but finite values of $D$.  It is important to note that
this wavefunction is constructed to display a specific long-range
behavior and that there is no direct control over the short-range
part.  The Hamiltonian with its nearest-neighbor terms may therefore
well favor a gap parameter $D$ which is shifted away from the value
that would correspond to the correct long-range behavior of the
groundstate.  Standard variational calculations provide no control
that would allow one to find out whether $D$, which determines the
long-range correlations, is over- or underestimated.  This is one of
the main disadvantages of the standard variational calculations.

We remedy this situation by optimizing the short-range correlations
independently from the variational wavefunction.  Since the
Hamiltonian only contains nearest-neighbor terms, we construct the
most general nearest-neighbor operator, which conserves the quantum
numbers for the spin and space symmetries:
\begin{equation}
{\cal A} = %
\alpha_0 + %
\alpha_1 \sum_{<i,j>,\sigma}%
              \left(\tilde{c}^\dagger_{i,\sigma}
                    \tilde{c}_{j,\sigma} + {\rm h.c.}\right) + %
\alpha_2 \sum_{<i,j>}{\bf S}_i \cdot {\bf S}_j + %
\alpha_3 \sum_{<i,j>}{1 \over 4} n_i n_j
\end{equation}
In the combined wavefunction
${\cal A}\left|\Psi\left(D\right)\right>$
we can adjust for the best short-range correlations through the choice
of the parameters $\alpha_i$.  The parameter $D$ now only controls the
long-range behavior for which it was designed.  By using an operator
${\cal A}$ with the same length scale as the Hamiltonian ${\cal H}$ we
arrive at a scheme, which is similar to a Lanczos iteration.  However,
we allow the parameters $\alpha_i$ to be adjusted independently of the
coupling strengths in the Hamiltonian.  Our approach can therefore be
regarded as a generalization of the Lanczos scheme.

In standard variational calculations the Rayleigh-Ritz principle (RR) is
used to find the best variational parameters.  The expectation value
$E_{\rm RR} = {\rm min}_\Psi %
\left<\Psi\right|{\cal H}\left|\Psi\right> / %
\left<\Psi|\Psi\right> %
$
is the lowest upper bound for the groundstate energy that can be
achieved with a given set of variational wavefunctions
$\left|\Psi\right>$.  Additionally the variance
$\sigma^2_{\cal H} = %
\left<{\cal H}^2\right> - \left<{\cal H}\right>^2
$
is small if a wavefunction is close to an eigenstate of the
Hamiltonian.  When approaching the groundstate, both of these values
should become smaller.  Although, this is a necessary condition, it is
not sufficient to distinguish from the case where one approaches an
excited eigenstate.

In our approach we use the same Rayleigh-Ritz principle but with the
addition of the generalized Lanczos operators (RRGL).  We arrive at
the expectation value
\begin{equation}
E_{\rm RRGL} = {\rm min}_{{\cal A},\Psi} %
{\left<\Psi\right|{\cal A}^*{\cal H}{\cal A}\left|\Psi\right> \over %
 \left<\Psi\right|{\cal A}^*{\cal A}\left|\Psi\right>} %
\end{equation}
which is still a lowest upper bound for the groundstate energy.  Also
the variance is again used to estimate the width of the energy
spectrum of the wavefunction.  Additionally to the standard
variational approach we can now compare the RR- to the
RRGL-wavefunction.  This allows us to judge whether the mean field
parameter $D$ in RR is indeed shifted to adjust for the short-range
behavior or whether the long-range correlations are maintained in the
groundstate.  Furthermore, we know from the Lanczos method used in
exact diagonalization that when one starts from a state with an energy
spectrum which is centered around an excited eigenstate, the first
iteration will already redistribute the weights towards lower
energies, such that the variance will first increase.  In this way we
use the variance as a unbiased indicator to judge the quality of the
RR-wavefunction.  Using these criteria the wavefunctions
Eq.~(\ref{wavefunction}), with d-wave order parameter, prove to be a
good choice consistent with the results from exact
diagonalization\cite{dagotto-riera93}.  In this work we will therefore
concentrate on the set of wavefunctions Eq.~(\ref{wavefunction}) as
our starting point.

In Fig.~\ref{energy} we show the energy for the RR wavefunction and
the improved RRGL wavefunction at quarter filling.  For 16 sites this
corresponds to 8 holes whereas for 50 sites we use 24 holes, so as to
have an even number of particles.  For the case of 8 holes on 16 sites
we can compare our results to the exact groundstate energy and we find
that the short-range correlations account for about 80 \% of the
missing correlation energy.  Fig.~\ref{variance} shows the square root
of the variance, {\it i.e.} the width of the energy spectrum.  This
value is considerably reduced for the RRGL wavefunction consistent
with a wavefunction which is closer to the groundstate.  For the 50
sites lattice there are no results for the exact groundstate energy
available and we have to estimate how the various quantities scale
with system size.  While the wavefunction scales with the size of the
system, the generalized Lanczos operators ${\cal A}$ always act on the
same length scale as the Hamiltonian.  The energy and $\sigma_{\cal
H}$ will therefore scale with the system size.  $\sigma_{\cal H}$ is
also the energy scale for the improvement $\Delta_E = E_{\rm RR} -
E_{\rm RRGL}$.  If the wavefunction has the right long-range behavior
then the operators ${\cal A}$ need to improve the correlations only on
the same length scale as the Hamiltonian and the ratio $\Delta_E /
\sigma_{\cal H}$ should remain constant.  Indeed we find this value to
be slightly bigger for the 50 sites lattice than for 16 sites.  This
again supports the observation that the wavefunction
Eq.~(\ref{wavefunction}) describes the correct long-range behavior of the
groundstate.  For the 50 sites lattice we investigated several
fillings, which all showed analogous results to the ones described
above.  Specifically we looked at the closed shell configurations of
8, 16, and 24 holes.

To investigate the superconductivity we measure the pair-pair
correlation function $C(R)$.  We find that for the 50 sites lattice
$C(R)$ is flat for the larger distances indicating that the finite
size effects are small.  We can therefore take $C_\infty = C(R_{\rm
max})$ as a measure for long-range order.  In the standard variational
approach $C_\infty$ is a monotonic function of $D$ and contains no
additional information.  With our new method we can now test how the
introduction of the operators ${\cal A}$ in the wavefunction affects
$C_\infty$.  If we start with too much long-range order the operators
${\cal A}$ will redistribute the weight in the correlation function
and suppress $C_\infty$.  On the other hand too small a value for
$C_\infty$ will be enhanced.  If we start with the correct long-range
order that corresponds to the groundstate, the operators ${\cal A}$
will only affect the short range part of $C(R)$ and $C_\infty$ will be
unaffected.  In that case we have effectively separated the short- and
long-range parts of the wavefunction.

We will illustrate this for the case of 8 holes on 50 sites
corresponding to a hole density of 0.16.  In Fig.~\ref{correlation}(a)
we show $C(R)$ for one value of $D=0.4t$ and $\mu=-0.8t$.  The solid
line corresponds to the RR-wavefunction.  We can see that the
long-range tail is well saturated.  For $J<J_c$ the long-range
correlations are suppressed, while for $J>J_c$ they are enhanced.
This is shown by the dashed lines.  For $D=0.4t$ we find $J_c \approx
1.0 t$.  We can now combine the data obtained for different gap
parameters $D$.  In Fig.~\ref{correlation}(b) we show $C_\infty$ as a
function of the coupling constant $J/t$.  The solid line again
corresponds to the RR-value while the dashed line shows the
suppression and enhancement for the RRGL-values.  The point where the
solid and dashed lines cross is the value $\widetilde{C}_\infty$ which
remains unchanged under iteration and we take this as the long-range
order $C_\infty$ of the groundstate.  For other values of the
variational parameter $D$ we repeat the same procedure and obtain the
points shown in Fig.~\ref{correlation}(b).  The error bars in $J/t$
indicate the region where the suppression or enhancement is within one
standard deviation.  We can thus map out $\widetilde{C}_\infty(J/t)$
for the groundstate and determine the critical value $J_c = (0.44
\pm 0.04)t$ for the onset of superconducting long-range order.  For
the other closed shell configurations of 16 and 24 holes we obtain
analogously the critical values $J_c = (0.62 \pm 0.08)t$ and $J_c =
(0.78 \pm 0.11)t$, respectively.  Using these points we construct the
phase diagram Fig.~\ref{phase-diagram}.  Ohta {\it et
al.}\cite{ohta-etal} observe a finite superconducting gap also at
lower values of $J/t$ for 16 and 18 sites systems.  However, they only
report results on open shells, in order to have $\bf k$ points at the
Fermi surface.  For open shells the Fermi sea is degenerate and the
d-wave condensate might be favored in order to remove this degeneracy.
Such finite size effects are reduced for the closed shell
configurations used in this work.

The wavefunctions used in this work describe a homogeneous electron
distribution for all variational parameters.  For large values of $J$,
the $t$-$J$ model exhibits phase
separation\cite{Emery-Kivelson-Lin,putikka-etal92}.  In that region of
the phase diagram, the groundstate is not well represented by the
homogeneous wavefunctions.  However, since the phase separated state
is a mixture of two homogeneous states --- the half filled Heisenberg
antiferromagnet (hole doping $\delta=0$) and a state with finite hole
doping --- we can use the Maxwell construction to obtain its energy.
For any fixed value of $J$ the energy of the {\it homogeneous}
wavefunctions will be a smooth function of the hole density $\delta$,
which we can describe by a polynomial in $\delta$.  In the region of
phase separation this polynomial will be curved downwards, so that a
combination of two homogeneous states (represented by a straight line)
will lower the energy.  The maximal hole density $\delta_c(J)$, which
lowers the energy between 0 and $\delta_c$ determines the phase
separation line as a function of $J$.  This line is also shown in
Fig.~\ref{phase-diagram}.  For the Maxwell construction we only use
the energies of the closed shell configurations (8, 16, and 24 holes),
for which we can expect the finite size effects to be minimal.  We can
then identify the region of d-wave superconducting long-range order
from close to quarter filling ($\delta=0.48$) down to a hole density
of $\delta=0.16$.  This is shown by the shaded region in the phase
diagram Fig.~\ref{phase-diagram}.

In conclusion we have presented a method which allows us to measure
the long-range behavior of the groundstate by separating long-range
from the short-range contributions.  This allows us to calculate the
long-range d-wave pair-pair correlation function, which is a direct
measure for the superconducting order parameter.  We identify the
region of d-wave superconducting long-range order down to a hole
density of $\delta=0.16$.  For smaller hole densities the calculations
will have to be extended to larger lattice sizes.  Our method extends
the results from the exact diagonalization, as it can be applied to
larger systems.

We would like to acknowledge useful discussions with S.~Barnes,
E.~Dagotto, G.~Blatter, N.~Bonesteel, C.~Gros, R.~Joynt, B.~Koltenbah,
T.~K.~Lee, H.~Tsunetsugu, and F.~C.~Zhang.  We thank the Swiss
Nationalfond for financial support.


\begin{figure}
\caption{%
Energy per site in units of $t$ as a function of coupling constant
$J/t$ for the $t$-$J$ model at quarter filling. $E_{\rm RR}$ is the
energy for the standard variational (Rayleigh-Ritz, RR) approach,
whereas $E_{\rm RRGL}$ shows the energy after optimizing the
short-range correlations with generalized Lanczos operators.  For 16
sites the values of the exact diagonalization from Dagotto {\it et
al.} (dotted line) are shown for comparison.  The lines are upper
bounds for the groundstate energy.  The statistical error is smaller
than the line width.}
\label{energy}
\end{figure}

\begin{figure}
\caption{%
Variance as a function of coupling constant $J/t$.  Shown is the
square root $\sigma_{\cal H}$ of the variance per site in units of
$t$.  This value measures the width of the energy spectrum and is a
measure for the quality of the wavefunction.  RR and RRGL correspond
to the same wavefunctions as in Fig. 1}
\label{variance}
\end{figure}

\begin{figure}
\caption{%
(a) d-wave pair-pair correlation function $C(R)$ as a function of
distance for a hole density of 0.16 and a gap parameter $D = 0.4t$.
The solid line corresponds to the raw RR wavefunction.  The dashed
lines show the long range correlation for the RRGL improvement for a
value of $J>J_0$ and $J<J_0$ respectively.  At the critical value
$J_0$ the long-range correlation is unchanged from the RR ansatz.  (b)
long-range d-wave correlation $C_\infty$ as a function of coupling
constant $J/t$.  The solid and dashed lines show $C_\infty$ for the
same gap parameter as in (a).  For other variational parameters only
the points $\widetilde{C}_\infty(J/t)$ where $C_\infty$ is unchanged
are shown.}
\label{correlation}
\end{figure}

\begin{figure}
\caption{%
Phase diagram of the $t$-$J$ model with parameters hole density
$\delta$ and coupling constant $J/t$.  The solid line shows the line
of phase separation.  The points indicate the onset of
superconductvity.  In the shaded region we observe a finite
superconducting long-range order.}
\label{phase-diagram}
\end{figure}

\end{document}